\documentclass[12pt]{iopart}
\usepackage{epsfig,rotate}

\begin{document}

\title[Study of the phase transition in the $3d$ Ising spin glass out of
equilibrium]{Study of the phase transition in the $3d$ Ising spin
  glass from out of equilibrium numerical simulations}

\author{S. P\'erez Gaviro \dag\ddag, Juan J. Ruiz-Lorenzo \S\ddag ~ and A.
  Taranc\'on \dag\ddag}

\address{\dag Departamento de F\'{\i}sica Te\'orica,
              Facultad de Ciencias,
              Universidad de Zaragoza, 50009 Zaragoza, Spain. }
\address{\S\ Departamento de F\'{\i}sica,
               Facultad de Ciencias,
               Universidad de Extremadura,
               06071 Badajoz, Spain.}
\address{\ddag Instituto de Biocomputaci\'on y F\'{\i}sica 
de Sistemas  Complejos (BIFI), Corona de Arag\'on 42, 50009 Zaragoza, Spain}
    
\eads{\mailto{spgaviro@unizar.es}, \mailto{ruiz@unex.es}, 
\mailto{tarancon@unizar.es}}

\date{today}

\begin{abstract}
  Using the decay of the out equilibrium spin-spin correlation function we
  compute the equilibrium Edward-Anderson order parameter in the three
  dimensional binary Ising spin glass in the spin glass phase.  We have
  checked that the Edward-Anderson order parameter computed from out of
  equilibrium numerical simulations follows with good precision the critical
  law as determined in experiments and in numerical studies at equilibrium
  (which allow us to estimate the $\beta$ critical exponent). 
  Finally we present a large time study  of the off-equilibrium
  fluctuation-dissipation relations and we find strong discrepancies (in the
  low temperature region)  between the numerical data and the droplet theory
  predictions and agreement with the predictions of the replica symmetry
  breaking theory.

\end{abstract}

\maketitle

\section{Introduction\label{sec:intro}}

The characterization, using numerical simulations, of the phase transition in
the three dimensional Ising spin glasses has been a challenging problem.
Recently a clear picture of the phase transition and good estimates of the
critical exponents have been obtained for both Gaussian and bimodal disorder
by working at equilibrium \cite{PRBNUM,KRATZ,JORG}.

However a characterization of the phase transition using out of equilibrium
techniques is still lacking (see reference \cite{3D} for a detailed
discussion).  In the first part of this paper we will address this problem
(simulating the bimodal disorder). In particular we will compute the order
parameter using out of equilibrium techniques \cite{DYN} and we will
characterize the transition using this observable. In addition we will
confront our data with previous estimates of the critical point and critical
exponents for this model (obtained from numerical simulations and from
experiments). The behavior of this observable will permit us to discard
(again) a Kosterlitz-Thouless like phase transition (as done in equilibrium
\cite{PRBNUM}, that we will refer in the following as $XY$-like scenario) for
the transition \cite{3D}. Moreover, we have studied the dependence of the
order parameter with the size of the system.  Hence, we will present on this
paper the first direct numerical computation of the Edwards-Anderson order
parameter in the three dimensional Ising spin glass (obtained out of
equilibrium).

This kind of study was performed in the past in four dimensions \cite{4D} (see
also \cite{DANIEL,YHT}) but is still lacking in three dimensions (the
interesting physical dimensions).

The second part of the paper is devoted to the study of the
fluctuation-dissipation theorem out of equilibrium. This kind of analysis have
attracted a large amount of work (analytical, numerical and experimental) in
the last years \cite{CUKU,FM,BCKP,FDT,FRARIE,FRANZ}.  

Using the results of reference \cite{FRANZ} and
assuming that the three dimensional Ising spin glass presents stochastic
stability (until now it has not been rigorously proved but there are
numerical evidences \cite{RSB}) one can relate the fluctuation-dissipation
curves with equilibrium properties and so, compute or measure the equilibrium
probability distribution of the overlap. This computation or measurement is
very important since it should discern between the different theoretical
approaches in competition, which try to describe the behavior of finite
dimensional spin glasses (e.g. the Replica Symmetry Breaking (RSB)
approach\cite{MPV,RSB} or the droplet model\cite{DROPLET}).

The goal of this (last) part of the paper is twofold. First, to check if the
order parameter computed in the first part of this paper matches well in the
fluctuation-dissipation (FD) curves. This is important since this value marks
the point in which the FD curve departs from its pseudo-equilibrium regime, and
the behavior of the curve from this departing point is a clear fingerprint
whether or not the system behaves following the RSB theory or the droplet model.

And the second goal is to study the finite time behavior (for really large
times) of the curves in order to see how the asymptotic form of the FD curves
is built up. This is important, since until now, the numerical simulations
\cite{FDT} and experiments \cite{HO_2002} show up a behavior compatible with
the Replica Symmetry Breaking  description  and
incompatible with droplet theory.  One can argue that the
curves reported in the literature \cite{FDT,HO_2002} are not asymptotic and
that the asymptotic curve is compatible with droplet theory and no compatible
with RSB.

Finally, we will report the conclusions.

\section{The model and Numerical simulations\label{sec:MOD}}

We have simulated a three dimensional system in a cubic lattice with
helicoidal boundary conditions of size $L$ and volume $V=L^3$. The Hamiltonian
is
\begin{equation}
{\cal H}=-\sum_{<i,j>} J_{ij} \sigma_i \sigma_j \,,
\end{equation}
where $<i,j>$ denotes the sum over the first nearest neighbors, $\sigma_i=\pm
1$ are Ising variables and $J_{ij}=\pm 1$ are quenched random variables
with a bimodal probability distribution with zero mean and unit variance.  We
have used the standard heat-bath algorithm (local dynamics) to simulate the
three-dimensional lattice.

We will introduce the observables measured in our work.
Firstly, the order parameter (the Edwards Anderson one) is defined as:
\begin{equation}
q_\mathrm{ EA}=\overline{\langle \sigma_i \rangle^2} \,,
\end{equation}
where, as usual, we use $\langle (\cdot \cdot\cdot) \rangle$ and 
$\overline{(\cdot \cdot\cdot)}$ to denote
thermal and quenched disorder average respectively.

In addition, 
the spin-spin correlation function has been computed using
\begin{equation}
C(t,t_w)=\frac{1}{V} \sum_{i=1}^V \sigma_i(t) \sigma_i(t_w) \,.
\end{equation}
We can obtain formally the order parameter from this correlation as the double limit:
\begin{equation}
q_\mathrm{ EA}=\lim_{t\to \infty} \lim_{t_w \to \infty}  C(t,t_w)\,.
\label{eq:limit}
\end{equation}
Notice that the order of the limit is crucial in obtaining the order parameter.
We will use this equation to extract $q_\mathrm{ EA}$ from the
out-of-equilibrium data.

We will study in the last part of the paper the finite time behavior of the
violation of the fluctuation-dissipation relation in the three dimensional
spin glass.  We will review shortly the main equation of the off-equilibrium
fluctuation-dissipation equations (see \cite{ME} for more details):
\begin{equation}
R(t_1,t_2)=\frac{1}{T} X(C(t_1,t_2))  \frac{\partial C(t_1,t_2)}{\partial t_2} \; ,
\end{equation}
where, $t_1>t_2$, $R(t_1,t_2)$ is the response of the system to the magnetic
field perturbation (i.e. the magnetic susceptibility of the system:
$R(t_1,t_2)=m(t_1,t_2)/h$) and $X(C)$ is the, in principle unknown, function
which controls the violation of the fluctuation-dissipation theorem.
Integrating this equation in $t_2$ and taking the perturbing field as
$h(t)=h\theta(t-t_w)$ we finally obtain (working 
in the linear-response region):
\begin{equation}
m(t) \simeq \beta h\int_{C(t,t_w)}^1 du ~X(u) \,.
\end{equation}

In the regime $t_1\gg t_2\gg 1$ we reach the equilibrium, and it is possible
to show that $C(t_1,t_2) \to q$. In addition $X(q) \to x(q) \equiv
\int_{q_\mathrm{min}}^q dq^\prime P(q^\prime)$, where $x(q)$ is the integral
of the probability distribution of the overlap at equilibrium \cite{MPV}. 
Hence, in this regime \cite{CUKU,FM,BCKP,FDT,FRARIE,FRANZ},
\begin{equation}
m(t) \simeq \beta h \int_{C(t,t_w)}^1 du~ x(u) \,.
\end{equation}

Furthermore, we can define
\begin{equation}
S(C) \equiv \int_{C(t,t_w)}^1 dq ~x(q) \,,
\end{equation}
so,
\begin{equation}
\frac{m(t) T}{h} \simeq S(C(t,t_w)) \,.
\end{equation}
Both, in droplet theory and RSB (see reference \cite{FDT2}, in particular its figure
10), $S(C)$ is the straight line $1-C$ for $C\in [q_\mathrm{EA},1]$. However,
for $C < q_\mathrm{EA}$ the behavior is very different: in the droplet theory
$S(C)$ is constant in this region and in RSB $S(C)$ is a growing function with
curvature. We recall that knowing the initial point, $S(C=0)$, we can compute
$q_\mathrm{EA}$ in the droplet theory as
\begin{equation}
q_\mathrm{ EA}^\mathrm{ droplet}=1-S(C=0) \,.
\end{equation}

This technique allows us to compute, taking the appropriate limit, the
equilibrium function $x(q)$.

Finally, we report that all the numerical simulations have been obtained with
the SUE machine~\cite{SUE}. This is a dedicated machine, designed for the
simulation of the three dimensional
Edwards-Anderson model with first neighbour couplings\cite{MPV}, the system
that is being studied in the present work. It consist of 12 identical boards.
Each single board is able to simulate $8$ different systems, updating all of
them at each clock cycle. SUE reaches an update speed of $217$ ps/spin with a
clock frequency of $48$ MHz. The on-board reprogrammability permits to change
in an easy way the lattice size, or even the update algorithm or the Hamiltonian.
The SUE machine is connected to a Host Computer running under Linux. SUE is in
charge of the update of the configurations, and the host computer is in charge
of measurements and analysis.  The main electronic devices of each SUE board
are the Altera family, that performs the update. Other devices store the spins
and couplings variables. One of the Alteras is devoted to generate random
numbers in a fast way (for more details, see Ref.~\cite{SUE}). Up our
knowledge, SUE has been the fastest dedicated machine in the simulation of the
three dimensional Edwards-Anderson model.

\section{Computation of the Edward-Anderson Order Parameter\label{sec:EA}}

\begin{figure}
\begin{center}
\leavevmode
\epsfig{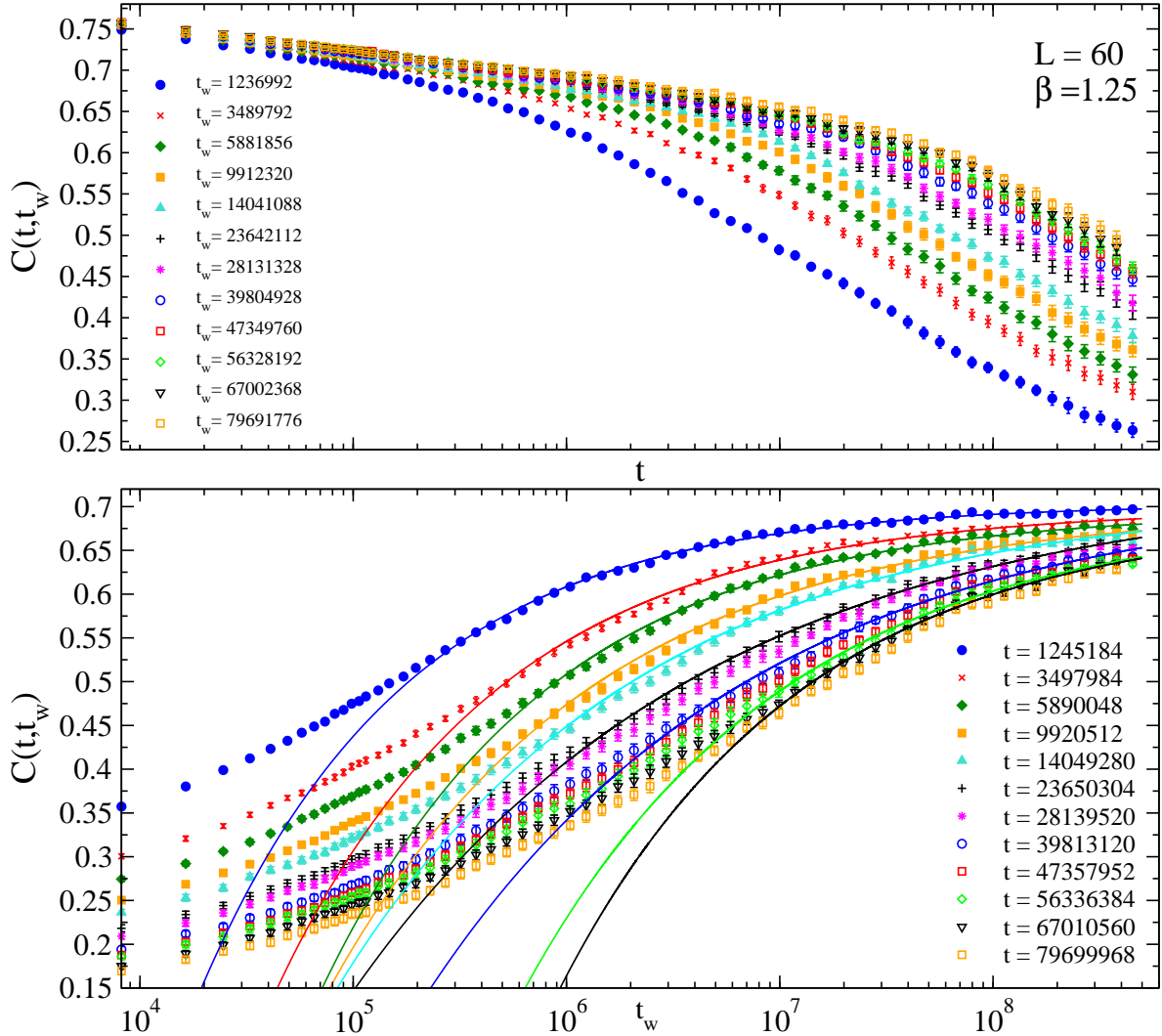}
\end{center}
\caption{Out of equilibrium spin-spin correlation function $C(t,t_w)$ computed
  for $L=60$ and $\beta=1.25$.  {\em {\bf Top}}: $C(t,t_w)$ versus time, $t$.
  {\em {\bf Bottom}}: $C(t,t_w)$ versus waiting time, $t_w$, obtained by
  studying figure in top for several fixed times $t$ in order to find the
  limit $t_w \to \infty$ behavior of $C(t,t_w)$. The continuous lines in the
  plot are the fits to equation (\ref{eq:fit}). Notice that for the curves
  with larger waiting time we have chosen to show  not all the fits to
  (\ref{eq:fit}) in order to present a clean figure (the quality of the fits is the same
  for all the waiting times).}
\label{Fig1}
\end{figure}

In order to compute the Edward-Anderson order parameter ($q_{\mathrm{EA}}$), 
we have
carried out several runs for two lattice sizes and different temperatures:
$\beta=1/T = 2.00$, $1.67$, $1.25$, $1.05$, $1.00$, $0.95$ and $0.91$ for
$L=30$; and $\beta = 2.00, 1.67, 1.25$ and $1.00$ for $L=60$ . For all of them
we have averaged over 58 samples. In figure (\ref{Fig1}) we report the curves
$C(t,t_w)$ as a function of time $t$.

\begin{figure}
\begin{center}
\leavevmode
\epsfig{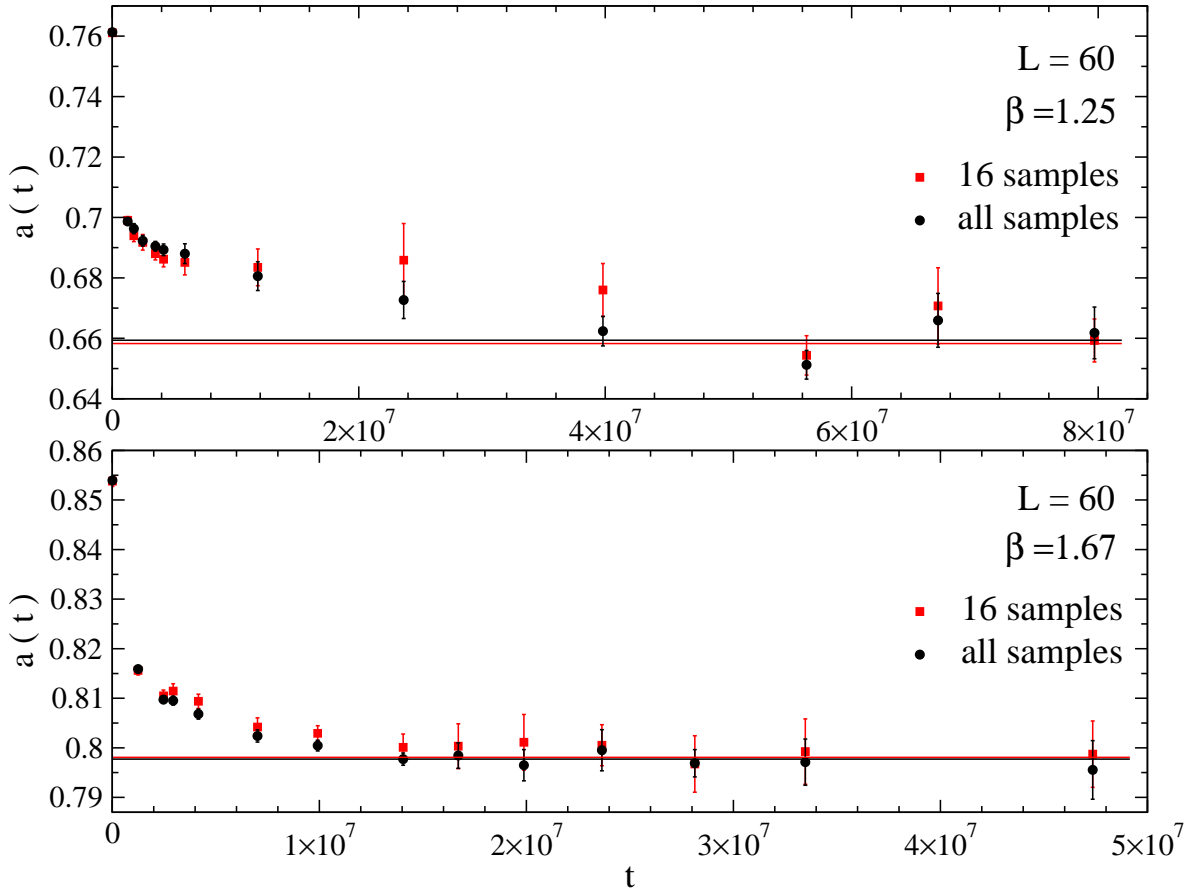}
\end{center}
\caption{Function $a(t)$ defined in equation (\ref{eq:fit})  computed for
  $L=60$ lattice size and $\beta=1.25$ (top) and  $\beta=1.67$ (bottom). 
Notice that the last points of the   curve can be fitted to a constant in the
two plots. In addition we have drawn the function $a(t)$ with only 16 samples
in order to show the dependence of the extrapolated value (i.e. the plateau)
on the number of samples.}
\label{Fig2}
\end{figure}

\begin{figure}[ht]
\begin{center}
\leavevmode
\epsfig{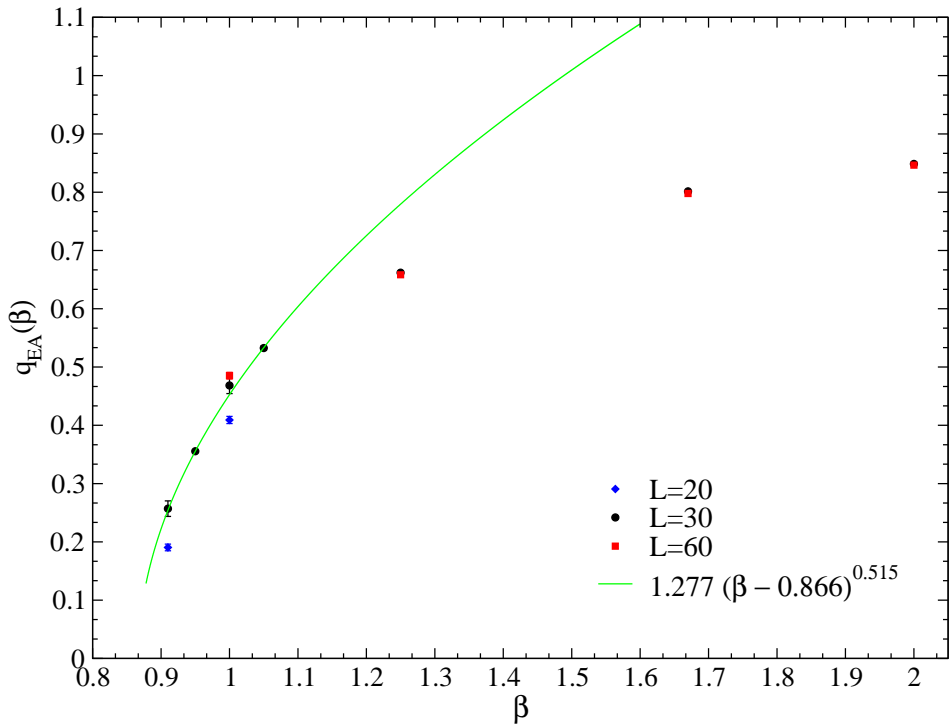}
\end{center}
\caption{$q_\mathrm{EA}^\mathrm{dyn}$ versus $\beta$ for three lattices sizes 
  $L=20, 30$ and $60$. The continuous line is the fit
  reported in the text.}
\label{Fig3}
\end{figure}

We have checked that the behavior of $C(t, t_w)$ for $t_w \gg 1$ follows with
high precision the behavior (as in higher dimensions, see \cite{4D,DANIEL};
this is just an Ansatz):
\begin{equation}
C(t, t_w)=a(t)+b(t) t_w^{-c(t)}\,,
\label{eq:fit}
\end{equation}
where $a(t)$ is related with the value of $q_{\mathrm{EA}}$.  In order to find
it out we have first obtained, from figure (\ref{Fig1}) top, the curves $C(t,
t_w)$ vs. $t_w$ for several fixed values of $t$ (typically, from $8192$ to
$\sim 3.7 \times 10^8$ Monte Carlo steps) (see figure (\ref{Fig1}) bottom). We
have fitted these curves to the functional form defined in (\ref{eq:fit})
obtaining in this way the behavior of $a(t)$ as function of $t$ (we show these
fits in figure (\ref{Fig1})) . From $a(t)$ and for $t \gg 1$, we can obtain
the value of $q_\mathrm{{EA}}$ (since asymptotically $a(t)$ must became
$q_\mathrm{EA}$). To achieve this aim, we have fitted the last points of
$a(t)$ versus $t$ to a constant function (since $a(t)$ shows a clear plateau,
see Fig.(\ref{Fig2})). In this way, we have implemented the double limit in
equation (\ref{eq:limit}).  The results obtained from these fits are shown in
Fig.(\ref{Fig3}).

We have checked that for $\beta>1.00$ the values for $q_\mathrm{EA}$ are the
same for both $L=30$ and $L=60$. In $\beta=1.00$ the difference is about 1.5
standard deviations. In addition we have run a $L=20$
lattice at $\beta=0.91 $ and $\beta=1.00$: these data show finite size effects
as expected since they lie near the critical point (see figure(\ref{Fig3})).

\section{Characterizing the Phase Transition\label{sec:PT}}

As we mentioned before, we have checked that the $q_{\mathrm{EA}}$, which we
have computed out of equilibrium, follows with good precision the critical law
of the order parameter
\begin{center}
\begin{equation}
q_\mathrm{{EA}}(\beta)=A (\beta-\beta_c)^{\beta_{q}}\,,
\end{equation}
\end{center}
where we have denoted $\beta_q$ the usual $\beta$ exponent of the order
parameter (in order to avoid confusion with the usual notation $\beta=1/T$)

By fitting  only the points closer to the critical one 
(satisfying $\beta<1.25$) we obtain
\begin{equation}
 \beta_c=0.866(2) \;\; \beta_q=0.52(9) \,,
\end{equation}
with a $\chi^2/\mathrm{d.o.f}=1.13$. This figures compare really well with the
numerical values obtained at equilibrium~\cite{PRBNUM}, namely:
$\beta_c=0.88(1)$ and $\beta_q=0.71(5)$. In particular the difference between
the two estimates of $\beta_q$ is 0.19(11), less than two standard
deviations.\footnote{Notice that in reference \cite{PRBNUM} corrections to
  scaling were taken into account. In our estimate there is no scaling
  corrections, hence our error are smaller than the error quoted in
  \cite{PRBNUM}: i.e. our error bars are underestimated.} \footnote{ See also
  ~\cite{Campbell} for a non Universality scenario: they reported
  $\beta_c=0.84(1)$.}

In addition, we can compare with experiments. In reference~\cite{PRBEXP} was
found $\beta_q=0.54(10)$\footnote{Note that both results in~\cite{PRBNUM}
  and~\cite{PRBEXP} come from different methods.} which is in a very good
agreement with our out equilibrium value.

We have also checked that  $q_{\mathrm{EA}}$ follows with good precision the
  critical law
\begin{center}
\begin{equation}
(q_{EA}(\beta))^{1/\beta_{q}}=A (\beta-\beta_c)\,,
\end{equation}
\end{center}
Again, we have only used in the fit the points with $\beta<1.25$ (critical
region).  Moreover we can fix $\beta_q$ to the experimental value, obtaining
again a compatible value with the equilibrium one: $\beta_c = 0.8603(6)(236)$,
where the first error is statistical and the second error comes from the error
of the experimental $\beta_q$. In addition, by fixing $\beta_q$ to the
numerical simulations value we obtain $\beta_c =0.820(3)(13)$, less than three
standard deviations from the numerical value.

All figures reported in this analysis are compatible with latest estimates of
the critical exponents. In reference \cite{KRATZ} $\beta_c=0.893(3)$ and
$\beta_q=0.723(25)$ were reported. In addition a diluted version of this model
was studied in \cite{JORG} and  $\beta_q=0.723(50)$ was reported.

Finally, we remark that our numerical results from both $\beta_c$ and
$\beta_q$ must suffer from the systematic error coming from the dependence of
$q_\mathrm{EA}$ with $L$ near the critical point (as shown the $L=20$ runs).
At $\beta=1.00$ we have three different values of the order parameter that
fit  to the law 
$$
q_\mathrm{EA}(L)= q_\mathrm{EA}(\infty)+\frac{b}{L^c} \,,
\label{eq:fit2}
$$
where $b$ and $c$ are constants.  This is the finite volume correction
equation which holds in the low temperature phase~\footnote{In reference
  \cite{XY} was checked that in the three dimensional Gaussian Ising spin
  glass the position of the maximum of the equilibrium probability
  distribution of the overlap follows this law with $c=1.5(4)$ by fitting $L
  \le 16$. Notice that in our case we are using $20 \le L \le 60$ data and we
  simulate the $\pm J $ model and that the $c$ exponent could depend on the
  temperature.  Notice that usually in equilibrium small lattices develop
  larger order parameter, however, in our dynamical approach we have found the
  opposite behavior.}. We have obtained $c=3.54$ and
$q_\mathrm{EA}(\infty)=0.49$ (notice that we are fitting three points to a
three parameter function) to be compared with $q_\mathrm{EA}(L=60)=0.485(6)$
and $q_\mathrm{EA}(L=30)=0.47(1)$.  At $\beta=0.91$ (the nearest value we have
to the critical point) we have only two points, that anyhow, we can try to fit
to equation (\ref{eq:fit2}) fixing $c=3.54$, obtaining
$q_\mathrm{EA}(\infty)=0.278$ (no error bars can be reported since, again, the
number of degrees of freedom in this fit is zero) to be compared with the
value of our largest lattice $q_\mathrm{EA}(L=60)=0.26(1)$, so this limited
analysis suggests that the $L=30$ lattice is asymptotic in its error bars in
the region $\beta\ge 0.91$. Hence, we are confident that our final estimates
of $\beta_c$ and $\beta_q$ should have small systematic error coming from
finite size effects.

We remark that testing the dependence of $q_\mathrm{ EA}$
with the lattice size, for large lattices (e.g. $L=60$) near the transition is
not accessible even using the SUE machine.

\section{Finite Time Effects in the Fluctuation-Dissipation relations\label{sec:FDT}}

We have performed several runs again with SUE machine, in a lattice of size
$L=60$ for different temperatures: $\beta = 1.25, 1.10, 1.05,
1.00$ and $0.95$.  We have used the following standard procedure. We let the
system evolve during a time $t_w$, just after this time, a field $h=0.03$ is
plugged, seeing the response of the system and recording the magnetization and
the correlation function. Then it is possible to extract the value of
$q_{EA}$, for the particular $\beta$ being analyzing at that moment, from the
point where the curve leaves the linear regime, that is, where $m T/h$ does
not follow the pseudo-equilibrium line $(1-C)/T$.

\begin{figure}
\begin{center}
\leavevmode
\epsfig{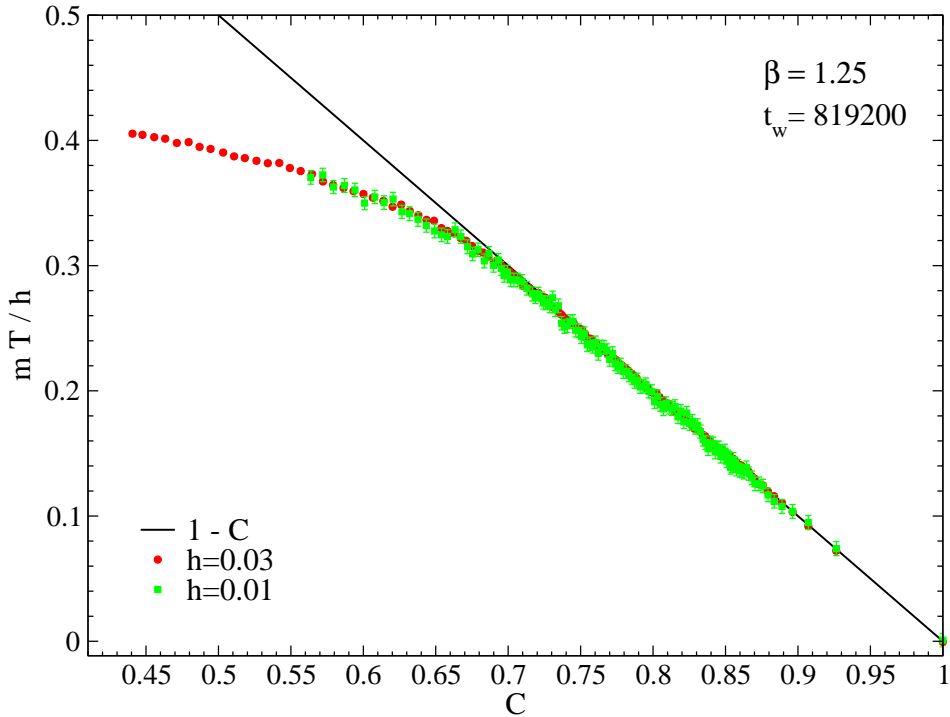}
\end{center}
\caption{Fluctuation-dissipation curve out of equilibrium for one of the
  lowest temperature simulated $\beta=1.25$, $L=60$ and one waiting time for
  two different perturbing magnetic fields, $h=0.01$ and $0.03$, in order to
  check linear-response.}
\label{FigF}
\end{figure}

The choice of the field strength applied to the system has not been arbitrary.
We need to stay in the linear-response region. We have checked this by
simulating different magnetic fields: $h=0.01, 0.03, 0.05$ and $0.10$. Finally
we have selected a safe value for $h$: $h=0.03$, which is a compromise between
large and small fields (notice that small fields induce strong noise in the
measures). In figure (\ref{FigF}) we have shown the FDT curve for a waiting
time and two perturbing magnetic fields ($h=0.01$ and 0.003) in order to test
that we are in the region in which linear-response holds. It is clear from
this figure that the curve, inside the error bars, is independent of the
perturbing magnetic field.

In the droplet model, the curve $X(C)$ departs horizontally from the straight
line $1-C$, the final value of the horizontal line being $m_\mathrm{ asyn}
T/h$ (i.e. $S(C=0)$), where $m_\mathrm{asyn}$ is the equilibrium value of the
magnetization in a field $h$ at the temperature $T$. Hence, measuring
$m_\mathrm{asyn}$ we can obtain the droplet theory estimate for the order
parameter as:
\begin{equation}
q_\mathrm{ EA}^\mathrm{ droplet}=1-\frac{m_\mathrm{asyn} T}{h}\,.
\end{equation}

\begin{figure}
\begin{center}
\leavevmode
\epsfig{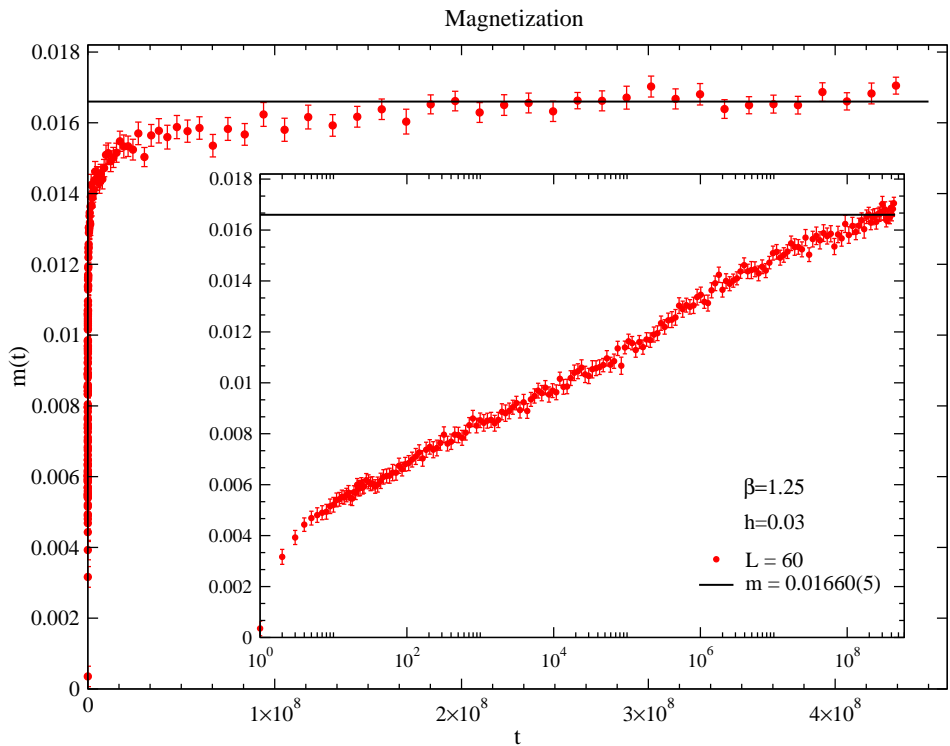}
\end{center}
\caption{Magnetization as a function of time for $\beta=1.25$, $L=60$ and
  $h=0.03$. We have plotted in the inset the main plot with logarithmic scale
  in the abscissas, in order to make sure we had found a plateau. We have
  marked in the main figure the plateau of the magnetization with a horizontal
  line.}
\label{Fig4}
\end{figure}

We will shown in this section plots corresponding to $\beta=1.25$ and $L=60$.
In order to obtain numerically $m_\mathrm{asyn}$ we have performed a very
large in-field numerical simulation recording the value of the magnetization
at the time $t$: $m(t)$. The asymptotic value is simply
$m_\mathrm{asyn}=m(\infty)$ (this observable shows really small dependence on
$L$ for the lattice sizes simulated in this paper). To avoid extrapolations we
have continued the run until the magnetization shows a plateau (this means
that the magnetization has reached its equilibrium value), and so we extract
the value of $m_\mathrm{asyn}$ by computing the position of this plateau.  For
instance, we show in figure (\ref{Fig4}) the magnetization as a function of
time for $\beta=1.25$ and $L=60$.

By computing the asymptotic value of the magnetization for different
temperatures, we obtain a reliable estimate for the order parameter in the
droplet theory. In Table \ref{Table_droplet} we report these values for the
droplet theory estimates and, in addition, we write the values for the order
parameter obtained in the first part of this paper, that we will denote in the
rest of the paper as $q_\mathrm{EA}^\mathrm{dyn}(\beta)$.

\begin{table}
\begin{center}
\begin{tabular}{c c c }
\hline
$\beta$ & $q_\mathrm{EA}^\mathrm{dyn}(\beta)$ & $q_\mathrm{
  EA}^\mathrm{droplet}$\\
\hline \hline
1.25    & 0.6583(34)  &  0.5573(13) \\ \hline
1.00    & 0.5071(31)  &  0.3957(17) \\ \hline 
0.95    & 0.3554(7)  &  0.3404(21) \\\hline
\end{tabular}
\caption{\label{Table_droplet}  $q_\mathrm{ EA}(\beta)$  for $L=60$ from 
  $C(t,t_w)$ (obtained in the first part of the paper) and assuming droplet
  theory from $m T/h$.  All the data showed in this table were obtained in a
  $L=60$ lattice except for dynamical $q_\mathrm{EA}$ at $\beta=0.95$ that was
  obtained simulating a $L=30$ lattice.}
\end{center}
\end{table}

We recall that the values of $q_\mathrm{EA}^\mathrm{dyn}(\beta)$ reported in
Table \ref{Table_droplet} have small  finite size effects (taking into account
their error bars) as checked in
figure (\ref{Fig3}). Moreover, we have
found strong discrepancies between $q_\mathrm{EA}^\mathrm{dyn}(\beta)$ and
$q_\mathrm{ EA}^\mathrm{droplet}$ for small temperatures.

We will describe in the rest of the paper our results for the violation of FDT
out of equilibrium.

\begin{figure}
\begin{center}
\leavevmode
\epsfig{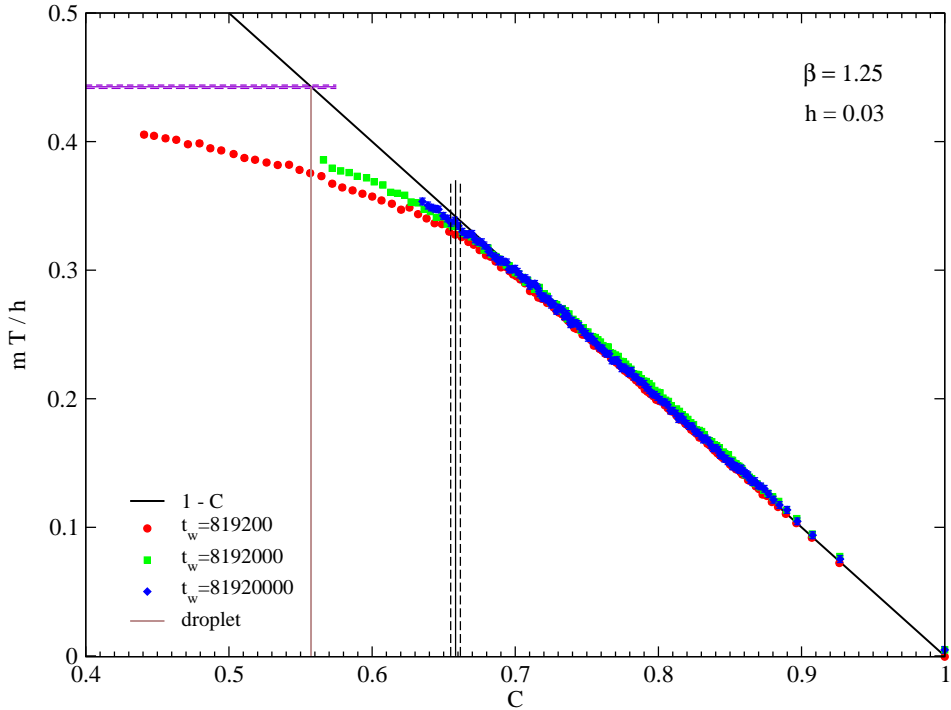}
\end{center}
\caption{Fluctuation-dissipation curve out of equilibrium for one of the
  lowest temperature simulated $\beta=1.25$, $L=60$ and three waiting times.
  We have marked using three vertical lines the interval in which lies
  $q_\mathrm{EA}^\mathrm{dyn}$ for this $\beta$ computed in the first part of
  the paper. In addition we have marked with three horizontal lines the value
  and the statistical error for $m_\mathrm{asyn} T/h$. Finally we have marked
  a vertical line with the droplet theory prediction for $q_{\mathrm{EA}}$
  (left part of the plot)}
\label{Fig5}
\end{figure}

In figure (\ref{Fig5}) we report the FD data out of equilibrium for one of
the lowest temperature simulated. We have shown a vertical band which marks the
our estimate of $q_\mathrm{EA}^\mathrm{dyn}$, a straight line $1-C$ to monitor
the departure of this linear behavior and  a horizontal band which marks
$m_\mathrm{asyn} T/h$ (see figure (\ref{Fig4})). In addition we have plotted
data from three different waiting times.

\begin{figure}
\begin{center}
\leavevmode
\epsfig{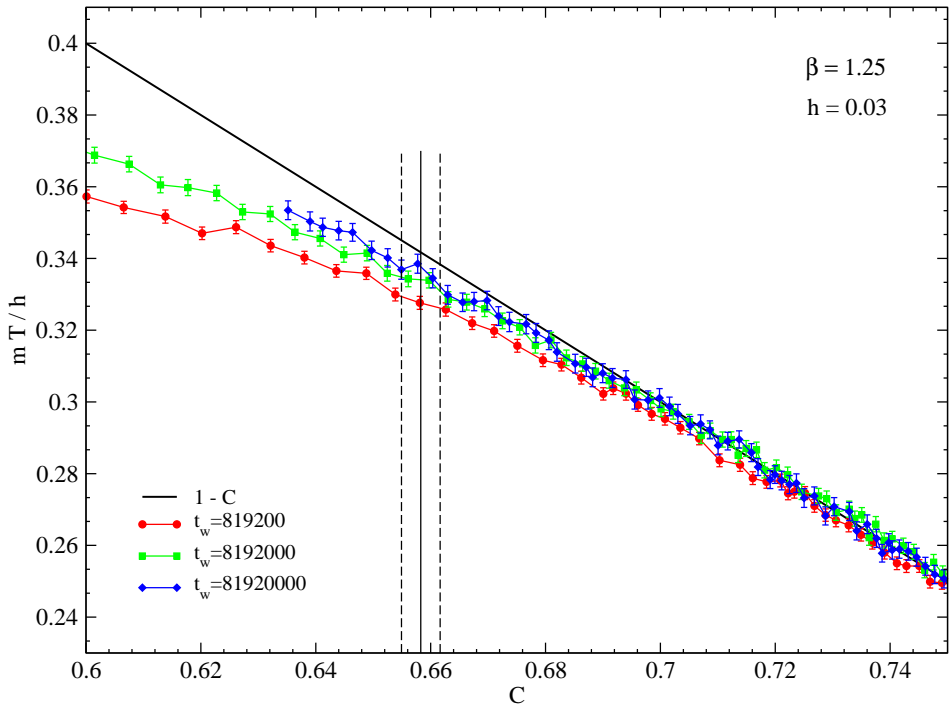}
\end{center}
\caption{Magnification of figure {\ref{Fig5}} showing up the region in which
  the FD curves depart from the straight line $1-C$. We have marked using
  three vertical lines the interval in which lies $q_\mathrm{EA}^\mathrm{dyn}$
  computed in the first part of the paper.}
\label{Fig6}
\end{figure}

Figure (\ref{Fig5}) shows that our estimate for $q_\mathrm{EA}^\mathrm{dyn}$
matches very well in the plot and marks the region in which the FD data starts
to depart from the linear behavior (for all the temperatures simulated). In
figure (\ref{Fig6}) we have drawn a magnification of this region. In addition,
in this figure one can see that the finite time effects in the building of the
asymptotic curve are small. Practically the two biggest waiting times are
compatible in the error (there is a factor ten in waiting time).  With the
state-of-the-art dedicated computed of the day it is impossible to simulate
larger waiting times. We can conclude from this figure that we are unable to
see dependence in waiting time for the two largest waiting times in the region
in which they depart from the linear behavior. The dependence on the waiting
time for larger times is smaller than our statistical errors.  From our
numerical data a droplet theory Fluctuation-dissipation asymptotic curve seems
unlikely.

\section{CONCLUSIONS\label{sec:CON}}

We have study numerically and out of equilibrium the three dimensional Ising
spin glass with bimodal disorder.

By computing the off equilibrium spin-spin correlation function we have been
able to extract the order parameter of the phase transition. The study of the
behavior of this order parameter with temperature permit us to compute the
critical temperature and the associated critical exponent: both figures 
compare very well with previous numerical simulations and experiments. We have
also discarded a $XY$-like scenario (we have found a non-vanishing order
parameter in the low temperature region). We have also monitored the dependence
of $q_\mathrm{EA}(\beta)$ with the lattice size in the low temperature region
for one $\beta$.

In the second part of the paper we have extracted the droplet prediction for
the order parameter by computing the asymptotic value of the susceptibility
($mT/h$). The droplet prediction compares (for all the $\beta$'s simulated) 
well with the order parameter
computed in the first part of the paper for high temperature (of course,
slightly below the critical temperature), but for lower temperatures the
comparison is bad.

Moreover the analysis (for larger waiting times) of the FD curves show a
behavior that can be described in the RSB theory and points out that the
droplet scenario seems unlikely (only a really small dependence on waiting
time, outside of the precision of this work, could build a final FD curve
compatible with the droplet theory). Moreover the point in which the numerical
data depart from the linear behavior compares well with the estimate obtaining
in the first part of this paper, supporting the RSB scenario.

\ack

This work has been partially supported by MEC (BFM2003-C08532, FISES2004-01399
and FIS2004-05073) and European Comission HPRN-CT-2002-00307. S. P\'erez
Gaviro is a D.G.A (Arag\'on Government) fellow.

\end{document}